\documentclass[preprint,12pt]{elsarticle} 
\usepackage{epsfig}
\usepackage{bm}

\begin{document}

\begin{frontmatter}

\title{Chaotic neuron clock}
\author{ A. Bershadskii$^{1,2}$ and Y. Ikegaya$^{3,4}$}

\address{$^1$ICAR - P.O. Box 31155, Jerusalem 91000, Israel and $^2$ICTP - Strada Costiera 11, I-34100 Trieste, Italy\\
$^3$Laboratory of Chemical Pharmacology, Graduate School of Pharmaceutical Sciences, 
University of Tokyo, Tokyo 113-0033, and \it $^4$Precursory Research
for Embryonic Science and Technology, Japan Science and Technology Agency, Kawaguchi 332-0012, Japan}

\begin{abstract}
A chaotic model of spontaneous (without external stimulus) neuron firing has been analyzed 
by mapping the irregular spiking time-series into telegraph signals. 
In this model the fundamental frequency of chaotic R\"{o}ssler attractor provides 
(with a period doubling) the strong periodic component of the generated irregular signal. 
The exponentially decaying broad-band part of the spectrum of the 
R\"{o}ssler attractor has been transformed by the threshold firing mechanism into 
a scaling tale. These results are compared with irregular spiking time-series obtained 
in vitro from a spontaneous activity of hippocampal (CA3) singular neurons 
(rat's brain slice culture). The comparison shows good agreement between the model and 
experimentally obtained spectra.

\end{abstract}
\begin{keyword}
Chaos \sep neuron  \sep spikes \sep spontaneous activity \\

\end{keyword}

\end{frontmatter}

\section{Introduction}

In order to work together the brain neurons have to make an adjustment of their rhythms. 
The main problem for this adjustment is the very noisy environment of the brain neurons. 
For pure periodic inner clocks this adjustment would be impossible due to the noise. 
Nature, however, has another option. This option is 
a chaotic clock. In chaotic attractors certain characteristic frequencies can be embedded by 
broad-band spectra, that makes them much more stable to the noise perturbations \cite{ra}. 
We will present empirical evidence supporting the proposition 
that the R\"{o}ssler chaotic attractor in combination with an appropriate threshold passage mechanism
could be used as a simple model of the spontaneous (without external stimulus) firing activity 
frequently observed in {\it in vitro} hippocampal neurons \cite{sas},\cite{taka}. 

The spontaneous firing activity should be more simple and self-consistent than a reaction of 
a neuron to the external stimuli. This can allow an analysis of the inner neuron clock in its 
"free-run" mode. Consideration of the most frequently firing neurons can also help in this direction. 
Besides its pure academic significance for studying the neurons' firing \cite{maso}
the spontaneous activity provides a considerable contribution to network 
development \cite{ps},\cite{zp}, information processing \cite{pare}-\cite{shu},
and behavioral responses \cite{brig},\cite{ott}. It is known that periodic spontaneous 
bursts of the activity can convey information about sensory stimuli \cite{ken},\cite{mac}. 
Spontaneous activity in brain slice 
preparations purely reflects the intrinsic properties of local circuits and individual 
neurons and hence allows for the
investigation of the internal dynamics of neuronal networks \cite{san}-\cite{sp}.\\

\begin{figure} \vspace{-2cm}\centering
\epsfig{width=.6\textwidth,file=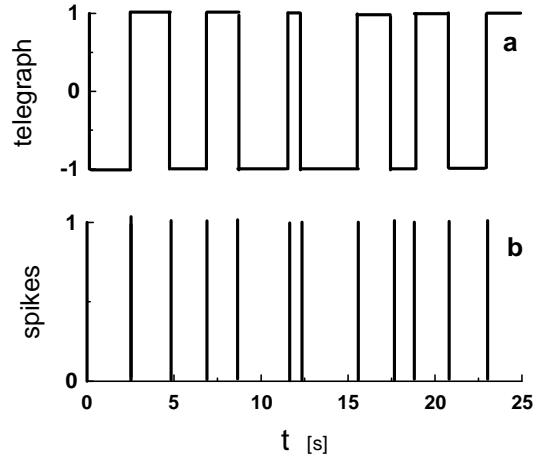} \vspace{-5cm}
\caption{Mapping of a spike train (figure 1b) into a telegraph signal (figure 1a). }
\end{figure}

   All types of information, which is received by sensory system, are encoded by nerve cells 
into sequences of pulses of similar shape (spikes) before they are transmitted to the 
brain. Brain neurons use such sequences as main instrument for intercells 
connection (both for tuning of their chaotic clocks and information-wise). 
The information is reflected in the time intervals between successive firings 
(interspike intervals of the action potential train, see Fig. 1b \cite{riek}). There need be no loss of 
information in principle when converting from dynamical amplitude information to spike trains 
\cite{sau} and the irregular spike sequences are the foundation of neural information 
processing. Although understanding of the origin of interspike intervals irregularity has important 
implications for elucidating the temporal components of the neuronal code and for treatment 
of such mental disorders as depression and schizophrenia, the problem is still very far from 
its solution (see, for instance, Ref. \cite{vs} and references therein). Motivation to study the 
hippocampus in relation to depression is based on the recently discovered evidences
of its deep involvement in this mental disorder. The
hippocampus is a significant part of a brain system responsible
for behavioral inhibition and attention, spatial
memory, and navigation. It is also well known that
spatial memory and navigation of the rats is closely related
to the rhythms of their moving activity. On the
other hand, the hippocampus of a human who has suffered
long-term clinical depression can be as much as 20\%
smaller than the hippocampus of someone who has never
been depressed \cite{brem}. One can speculate that in the case of
depression the chaotic neuron clocks can be
broken in a significant part of the brain neurons. That
can result in certain decoherence in different parts of
the brain.

\begin{figure} \vspace{-2cm}\centering
\epsfig{width=.6\textwidth,file=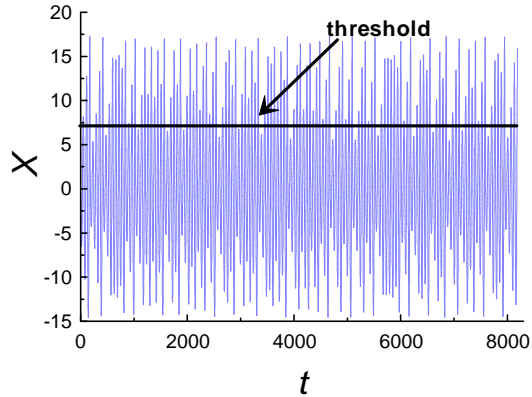} \vspace{-5.5cm}
\caption{X-component fluctuations 
of a chaotic solution of the R\"{o}ssler system Eq. (1) ($a=0.15,~ b=0.20,~ c=10.0$). }
\end{figure}

Treatment of the spiky signals by analytical methods also presents a difficult problem. 
In order to apply the Fourier transform method one can use different mappings: 
mapping of the finite spikes into Dirac delta functions or into telegraph signals, for instance (cf. Ref. \cite{lt}).  
For a single neuron firing the amplitude of the spikes are almost identical to each other 
and the neural information is coded in the length of the interspike intervals and their positions on the time axis \cite{riek},\cite{sau}, therefore it is the most direct way to map the spike train 
into a telegraph time signal, which has values -1 from one side of a spike and values +1 from 
another side of the spike with a chosen time-scale resolution. An example of such mapping is given in figure 1. 
While the information coding is here the same as for the corresponding spike train, the Fourier 
transform methods are quite applicable to analysis of the telegraph time-series \cite{luk}. 

\section{A spontaneous firing model: the R\"{o}ssler system}

 Nerve cells are surrounded by a membrane that 
allows some ions to pass through while it blocks the 
passage of other ions. When a neuron is not sending a 
signal it is said to be "at rest". At rest neurons exhibit very small conductance of sodium ions and
slightly larger potassium conductance against a high concentration of intracellular potassium ions. 
The resting value of {\it membrane} electrochemical potential - $P$
(the voltage difference across the neural membrane) of a neuron is about -70mV. If some event 
(a stimulus) causes the resting potential to move toward 0mV and the depolarization reaches about -55mV 
(a "normal" threshold) a neuron will fire an {\it action} potential. 
The action potential is an explosive release of charge across plasma membrane and its surrounding that is 
created by a depolarizing current. If the neuron does not reach this critical threshold level, 
then no action potential will fire. Also, when the threshold level is 
reached, an action potential of a {\it fixed} size will always fire (for any given neuron the size of 
the action potential is always the same). 

\begin{figure} \vspace{-2cm}\centering
\epsfig{width=.6\textwidth,file=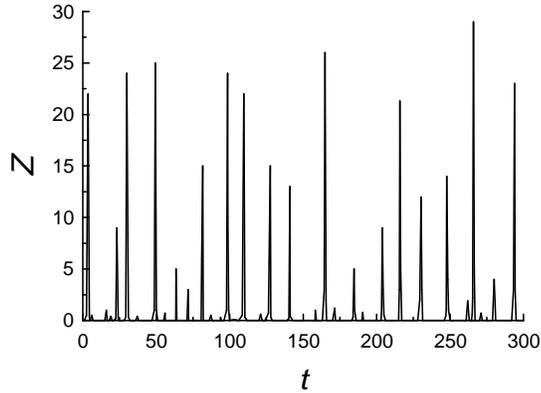} \vspace{-6cm}
\caption{Z-component fluctuations 
of a chaotic solution of the R\"{o}ssler system (the parameters are the same as in Fig. 2). }
\end{figure}

Recent reconstructions of a driver of the {\it membrane} potential using the 
neuron spike trains indicate the R\"{o}ssler oscillator as the most probable (and simple) 
candidate (see, for instance, Refs.\cite{cas}-\cite{per}). 
Figure 2 shows as example the x-component fluctuations 
of a chaotic solution of the R\"{o}ssler system \cite{ross} 

$$
\frac{dx}{dt} = -(y + z);~~  \frac{dy}{dt} = x + a y;~~  \frac{dz}{dt} = b + x z - c z \eqno{(1)}
$$      
where a, b and c are parameters. At certain values of the parameters a,b and c the $z$-component 
of the R\"{o}ssler system is a {\it spiky} time series Fig. 3 (see also Refs. \cite{llg},\cite{lgl}). 

It can be shown that the R\"{o}ssler system and the 
well known Hindmarsh-Rose model \cite{hr} of neurons are subsystems of the same differential model with 
a spiky component \cite{lgl}. Previously the 'spiky' component of such models was interpreted and studied 
as a simulation of a neuronal {\it output}. For the {\it spontaneous} neuron firing 
(without external stimulus), however, we suggest to reverse the approach and consider the spiky variable 
as the main component of the electrical {\it input} (which naturally should have 
a 'spiky' character, see above) to the neuron under consideration. For any given neuron the height of the 
spikes, which the neuron generates, is about the same. However, the heights of the spikes generated 
by different neurons are different. Also the signals coming from different neurons to the neuron under 
consideration have to go through the electrochemical passes with different properties. 
Therefore, the spiky $z$-time-series (Fig. 3) can naturally represent 
a multineuron signal, which can be considered as a spontaneous input for the neuron under consideration. 
\begin{figure} \vspace{-2cm}\centering
\epsfig{width=.6\textwidth,file=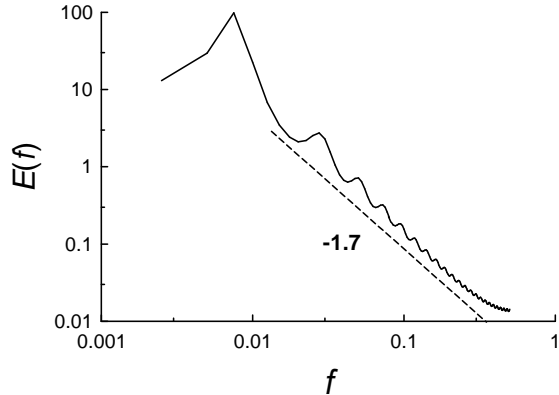} \vspace{-5cm}
\caption{Spectrum of the telegraph signal corresponding to the spike train generated by 
the x-component fluctuations overcoming the threshold $x=7$. 
The dashed straight line indicates a power law Eq. (3) in the log-log scales. }
\end{figure}
If we use the usual interpretation 
of the $x$-component as a driver of the membrane potential $P(x)$ and the $y$-component as that taking into 
account the transport of ions across the membrane through the ion channels \cite{hr}, 
then addition of the spiky z-component (representing in present model the multineuron spontaneous input) in the dynamical equation for x-component is similar 
to the addition of an external input-component to the dynamical x-equation in the Hindmarsh-Rose model. 
Then, the quadratic nonlinearity in the third equation of the system Eq. (1) can be interpreted as a simple
(in the Taylor expansion terms) feedback of the neuron to the main component of the neuronal input. 
This model with the strong nonlinear feedback can 
be relevant to the most active neurons of a spontaneously active brain (see below results of 
an {\it in vitro} experiment with a spontaneous brain activity). The details of the function $P(x)$ is not 
significant for the threshold firing process, what really matters is that the membrane potential 
function $P(x)$ reaches its firing value when (and only when) its argument $x$ crosses certain threshold 
from below. In this simple model the driving variable $x$ may overcome its threshold value (Fig. 2) 
due to the deterministic (chaotic) spontaneous stimulus. Let us consider an output spike signal resulting from 
overcoming a threshold value $x=7$, for instance. Fig. 4
shows spectrum of the telegraph signal corresponding to the spike train. 
\begin{figure} \vspace{-1cm}\centering
\epsfig{width=.6\textwidth,file=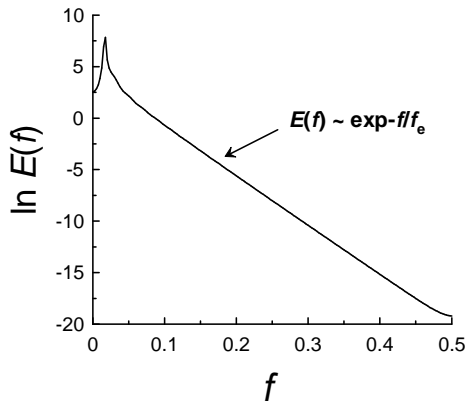} \vspace{-5cm}
\caption{Spectrum of the x-component fluctuations shown in Fig. 2. We used the semi-log 
axes in order to indicate exponential decay of the spectrum.
 }
\end{figure}
In order to understand what is going on here we show in figure 5 spectrum of the x-component itself. 
The semi-log scales are used in these figures in order to indicate exponential decay 
in the spectra (in the semi-log scales this decay corresponds to a straight line):
$$
E(f) \sim e^{-f/f_e}        \eqno{(2)}
$$
While the low-frequency peak in the spectrum corresponds to the fundamental frequency, $f_0$, of the 
R\"{o}ssler chaotic attractor, the rate of the exponentional decay (the slope of the 
straight line in Fig. 5) provides us with and additional characteristic frequency $f_e$ Eq. (2). 
\begin{figure} \vspace{-2cm}\centering
\epsfig{width=.6\textwidth,file=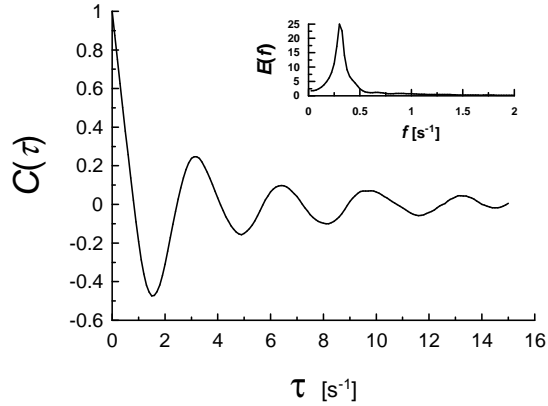} \vspace{-5cm}
\caption{Autocorrelation function for the telegraph signal corresponding to the cell-21 (800 spikes). 
The insert shows corresponding spectrum. }
\end{figure}
It should be noted that for a wide class of deterministic systems a broad-band spectrum 
with {\it exponential} decay is a generic feature of their {\it chaotic} solutions 
Refs. \cite{sig}-\cite{fm}. 

Thus R\"{o}ssler chaotic attractor has two clocks: fundamental with frequency $f_0$ and 
decaying with frequency $f_e$. If one compares Fig. 4 and Fig. 5 one can see that 
the fundamental clock survives the threshold crossing (with a period doubling, see also Fig. 9). 
The decaying clock, however, does not survive the threshold crossing: the exponential decay in 
Fig. 5 has been transformed into a scaling (power law) decay in Fig. 4
$$
E(f) \sim f^{-\alpha}        \eqno{(3)}
$$
(with $\alpha \simeq 1.7$), which has no characteristic frequency (scale invariance). \\

In the above presented model the fundamental frequency of chaotic R\"{o}ssler
attractor provides (with a period doubling) the strong periodic component
of the generated irregular signal. This periodic component can be utilized by Nature 
as a chaotic clock of the spontaneous neuron firing (see next section). 
The exponentially decaying broad-band part of the spectrum of the R\"{o}ssler attractor 
has been transformed by the threshold firing mechanism into a scaling tale. To understand 
mechanism of this transformation is rather difficult problem since the mechanism of generation 
of the exponential spectrum by the chaotic systems is itself an unsolved problem (see, 
for instance, Refs. \cite{sig}-\cite{fm}). The scaling exponent value '-1.7' is 
not sensitive to a reasonable variation of the threshold value ($\sim 20$\%) and even 
to Gaussian fluctuations of the threshold value. Therefore, it is not just a coincidence 
that the scaling law in the R\"{o}ssler case agrees with results of the {\it in vitro} experiment 
reported in the next section (cf. also \cite{luk},\cite{grig}). The power-law spectrum Eq. (3) 
with the exponent $\alpha \simeq 1.7$ is an indication 
of a strongly non-ergodic system, with perennial aging, or maybe that an infinitely aged correlation 
function is predominant due to noise or truncations. Moreover, the {\it in vitro} data, which we 
used for comparison (see next Section) seem to fit a renewal
hypothesis, compatible with this scaling exponent, and the
theory of a recent Ref. \cite{all} (see also Ref. \cite{ref}). 
\begin{figure} \vspace{-2cm}\centering
\epsfig{width=.6\textwidth,file=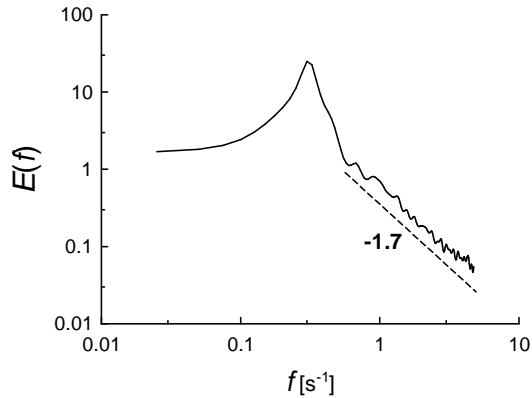} \vspace{-5cm}
\caption{Spectrum of the telegraph signal corresponding to the cell-21 (800 spikes) in log-log scales. 
The dashed straight line indicates a power law Eq. 4: $E(f) \sim f^{-1.7}$.  }
\end{figure}
\begin{figure} \vspace{-1cm}\centering
\epsfig{width=.6\textwidth,file=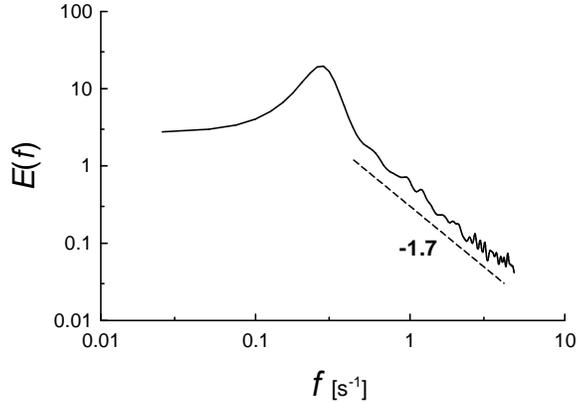} \vspace{-5cm}
\caption{As in Fig. 7 but for cell-25 (D-006, 692 spikes). }
\end{figure}

\section{In vitro spontaneous brain activity}

In order to compare this simple model consideration with the experimental data 
we have analyzed spike trains obtained {\it in vitro} from a spontaneous activity 
in CA3 hippocampal slice culture of a Wistar/ST rat (the raw data and the detail description of the experiment 
can be found online at http://hippocampus.jp/data and in Refs. \cite{sas},\cite{taka}), In the {\it in vitro} experiment a functional imaging technique with multicell loading of the calcium 
fluorophore was used in order to obtain the spike trains of spontaneously active singular neurons 
(hippocampal pyramidal cells) in the absence of external input. 
In this experiment different levels of activity were 
observed for different neurons \cite{sas},\cite{taka}. 
We take for our analysis the two most active neurons (http://hippocampus.jp/data - Data-006, 
cell-21, with 800 spikes in the time-series; and a cell-25, with 692 spikes). 
The spike trains were mapped to telegraph signals as it is described above. Figure 6 shows the 
autocorrelation function 
for the telegraph signal corresponding to the cell-21 (800 spikes). Insert in the Fig. 6 shows 
the corresponding spectrum. Both the correlation function and the spectrum provide a clear indication 
of a strong periodic component in the irregular signal (the oscillations in the correlation function and 
the peak in the spectrum). The periodic component can be seen at frequency $f_0 \simeq 0.3$Hz. 
Figure 7 shows the spectrum in log-log scales. One can see that at high frequencies the spectrum 
exhibits a scaling behavior Eq. (3) (power law: $E(f) \sim f^{-1.7}$, as indicated by the dashed straight line). 
The real power law can be more pronounced but under the experimental conditions individual spikes 
emitted at firing rates higher than 5Hz were experimentally inseparable \cite{sas},\cite{taka}. 
Figure 8 shows spectrum of the telegraph signals corresponding to the spike train obtained 
for the cell-25 (D-006, 692 spikes). The spectrum is rather similar to the spectrum shown in Fig. 7 (
for cell-21). The more broad peak in Fig. 8 can be related to the poorer statistics for the cell-25 
(number of spikes 692) in comparison with cell-21 (number of spikes 800). 
\begin{figure} \vspace{-2cm}\centering
\epsfig{width=.6\textwidth,file=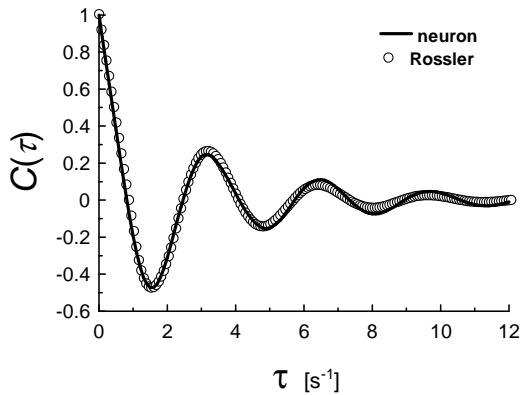} \vspace{-5cm}
\caption{Autocorrelation functions for the telegraph signals corresponding to the cell-21 
and to the spike train generated by  the R\"{o}ssler attractor fluctuations overcoming the threshold $x=7$ (circles). In order to make the autocorrelation functions comparable a rescaling has been made 
for the R\"{o}ssler autocorrelation function.  
 }
\end{figure}
One can compare Figs. 7 and 8 with Fig. 4 to see very good reproduction of the main spectral 
properties. Figure 9 shows also a superposition of the autocorrelation functions for the telegraph signals corresponding to the cell-21 (the solid line) and to the spike train generated by 
the the R\"{o}ssler attractor fluctuations overcoming the threshold $x=7$ (circles). In order to make 
the autocorrelation functions comparable a rescaling has been made for the R\"{o}ssler autocorrelation function:
the scaling coefficient is equal to 0.024 (the time step size for the R\"{o}ssler solution was 0.01). 
As in the case of the spectra the comparison of the autocorrelation functions 
is not sensitive to a reasonable variation of the threshold value (~20\%) and even to Gaussian fluctuations of the threshold value. 

\section{Conclusion}

Simple R\"{o}ssler system has properties, which allow to use it as an adequate model of {\it spontaneous} 
neuron activity (that includes a multineuron {\it input} represented by the spiky z-component of the 
R\"{o}ssler attractor, Fig. 3, and a strong nonlinear feedback). A mapping of the spike time series into telegraph signals (that preserves completely intact the 'frequency modulated' information of the spike series) allows to compare spectral properties of the model threshold firing (Fig. 4) with analogous firing of the in vitro brain neurons spontaneous activity (Figs. 7 and 8). This investigation can be considered as an additional confirmation and specification of the already existing evidences (Refs. \cite{cas}-\cite{per}) that the R\"{o}ssler system can be a driver of spontaneous neuron firing. On the other hand, it is known that the spontaneous firing of brain neurons accounts for about 80\% of the metabolic energy consumed by the brain \cite{rai}. Therefore the spontaneous 
neural activity should have a very significant neurobiological function (cf. Introduction). 

   Spontaneous neural activity with multineuron nonlinear interactions has been studied mainly 
in neuronal networks (see, for instance, Refs. \cite{sas}-\cite{sp}). In present simple model a multineuron 
input and a strong nonlinear feedback are simulated in the frames of a single self-consistent low-dimensional 
system with a chaotic solution. On the other hand, we do not know any study where a fully chaotic 
system could make anomalous behaviors emerge, so it seems that studying threshold passage is
enough to have that effect. 

Of course, in vivo neuron signals can be much more complex. The relatively 
simple firing of the in vitro spontaneously active hippocampal neurons can be used in order 
to reveal the underlying neuron dynamics.

\section{Acknowledgments}
We thank Takahashi N., Sasaki T., Usami A., and Matsuki N. for help in obtaining the data. 
We also thank Greenberg A. for help in computing.

\end{document}